\documentstyle[floats,twocolumn,aps,psfig,prl]{revtex}

\begin{document}
\draft

\twocolumn[\hsize\textwidth\columnwidth\hsize\csname @twocolumnfalse\endcsname

\title{Field-induced metal-insulator transition in a
two-dimensional organic superconductor}
\author{J. Wosnitza$^1$, S. Wanka$^1$, J. Hagel$^1$, H. v. L\"ohneysen$^1$,
J. S. Qualls$^2$\cite{address}, J. S. Brooks$^2$, E. Balthes$^3$,
J. A. Schlueter$^4$, U.~Geiser$^4$, J. Mohtasham$^5$,
R.~W.~Winter$^5$, and G. L. Gard$^5$}
\address{$^1$Physikalisches Institut, Universit\"at Karlsruhe, D-76128
 Karlsruhe, Germany\\
 $^2$National High Magnetic Field Laboratory, Florida State University,
 Tallahassee, Florida 32306\\
 $^3$Grenoble High Magnetic Field Laboratory, MPI-Festk\"orperforschung
 and C.N.R.S., F-38042 Grenoble, France\\
 $^4$Chemistry and Materials Science Divisions,
  Argonne National Laboratory, Argonne, Illinois 60439\\
 $^5$Department of Chemistry, Portland State University,
  Portland, Oregon 97207}
\date{\today}
\maketitle

\begin{abstract}
The quasi-two-dimensional organic superconductor
$\beta^{''}$-(BEDT-TTF)$_2$SF$_5$CH$_2$CF$_2$SO$_3$ ($T_c \approx 4.4$\,K)
shows very strong Shubnikov--de Haas (SdH) oscillations which are
superimposed on a highly anomalous steady background magnetoresistance,
$R_b$. Comparison with de Haas--van Alphen oscillations allow a
reliable estimate of $R_b$ which is crucial for the correct extraction
of the SdH signal. At low temperatures and high magnetic fields
insulating behavior evolves. The magnetoresistance data
violate Kohler's rule, i.e., cannot be described within the
framework of semiclassical transport theory, but converge onto a
universal curve appropriate for dynamical scaling at a
metal-insulator transition.
\end{abstract}


\vskip2pc]

The electrical transport in metals can usually be described by the
coherent motion of electrons in Bloch states with well-defined
wave vectors. A common approach to
this problem is the Boltzmann transport theory which works well for
most metals and semiconductors. There are, however, a number of
cases where a more complex transport mechanism is
involved and where the simple approach fails \cite{mck98,mck98b}.
Prominent examples are the cuprate superconductors \cite{and97} and
organic metals \cite{str94} which reveal unusual normal-state
properties. A central issue for these layered materials is whether
the electronic conduction can be described by the coherent motion
of Bloch electrons with well-defined wave vectors or whether the
interlayer transport is caused by an incoherent diffusive motion
of the electrons between the layers.

With the assumption of a constant scattering time $\tau_s$ for all
charge carriers the semiclassical transport theory predicts a
universal temperature and field dependence of the magnetoresistance
which can be described as $R(B,T)/R(0,T) = f(B/R(0,T))$, where
$f(x)$ is a universal function. This is known as Kohler's rule
which holds for many metals regardless of the Fermi-surface
topology \cite{pip89}. Furthermore, for $B$ parallel to the
current, {\it no} magnetoresistance is expected semiclassically.
Deviations from this behavior are known
to occur for the interlayer transport in some organic conductors
\cite{mck98b,str94}. This was taken as an indication for incoherent
transport. Other evidence for the failure of conventional
transport theory is the very large low-temperature
normal-state resistivity (a few $\Omega$cm)
which would correspond to mean-free paths much shorter than
interatomic distances.

For the quasi-two-dimensional (2D) organic metals one can assume that
the interlayer transport is caused by uncorrelated tunneling
events between the layers \cite{mck98}. Thereby the transport is
incoherent because the electrons are scattered many times within
the layer before a tunneling event takes place. This may occur when
the time it takes for an electron to hop between the layers is much
larger than $\tau_s$, i.e., $\hbar/t_c \gg \tau_s$, where $t_c$ is the
interlayer hopping integral. In case the intralayer momentum is
conserved during the tunneling process and an interference between
the wave functions on adjacent layers is possible, McKenzie and Moses
\cite{mck98} showed that certain metallic properties persist even
though no three-dimensional (3D) Fermi surface would exist.

A potential candidate which might fit into the above scenario is the
2D organic superconductor
$\beta^{''}$-(BEDT-TTF)$_2$SF$_5$CH$_2$CF$_2$SO$_3$ ($T_c = 4.4$~K)
\cite{wan98}, where BEDT-TTF
stands for bis\-ethylene\-dithio-tetrathiafulvalene.
The Fermi surface has been mapped out by de Haas--van Alphen (dHvA)
\cite{wos98}, Shubnikov--de Haas (SdH) \cite{bec98}, and angular-dependent
magnetoresistance oscillations (AMROs) \cite{wos99}. High-field dHvA
measurements proved the existence of an ideal 2D Fermi surface
\cite{wos00}. In line with an incoherent transport mechanism \cite{mck98}
neither beats in the magnetic quantum oscillations nor a peak in
the AMROs for field parallel to the layers was observed.
Another not-explained phenomenon is the peak in
$R(B)$ at low fields and low temperatures \cite{su99}.
At zero field, the material seems to be close to an insulating
phase, since replacing CH$_2$CF$_2$ in the anion with
CH$_2$ yields an insulator \cite{war00}.

Although the measured dHvA oscillations of
$\beta^{''}$-(BEDT-TTF)$_2$SF$_5$CH$_2$CF$_2$SO$_3$
\cite{wos00} could quantitatively be
understood by a 2D theory \cite{sho84,har96}, the SdH oscillations
seemed to show strong deviations in the field and temperature
dependence from the usually expected behavior \cite{wos99,zuo99}.
Similar observations have been reported for the SdH oscillations in
other organic metals \cite{lau95,san96,bal96,har98}.
A principal problem inherent to transport data is the
correct extraction of the SdH signal, which is
given by the relative conductance oscillations $\Delta \sigma/\sigma
= \sigma/\sigma_b - 1$, with $\sigma_b$ the steady part of the
relevant conductance of the band which is responsible for the
oscillations \cite{ada59}. Only for $\Delta \sigma/\sigma$ the
framework of the Lifshitz--Kosevich (LK) theory can be applied
\cite{sho84}. In general, a tensor inversion from the measured
longitudinal and transverse resistances is necessary to obtain
$\sigma_b$. In the present case, however, the Hall component of
the resistivity tensor is negligible and for the used field
configuration ($B$ is applied {\it parallel} to the interlayer-transport
current) it should be even exactly zero. Therefore, the SdH signal
is given by $\Delta \sigma/\sigma = R_b/R - 1$, where the main
task is reduced to the reliable determination of $R_b =
1/\sigma_b$ out of the measured resistance $R$. In the following
we specify a method how this can be reasonably performed.
The extracted $R_b$ reveals a field-induced insulating behavior,
which cannot be described by the semiclassical transport theory
but can be scaled onto a universal curve.
The observed dynamical scaling suggests a scenario of this
metal-insulator transition as a quantum phase transition.

\begin{figure}
  \centerline{\psfig{file=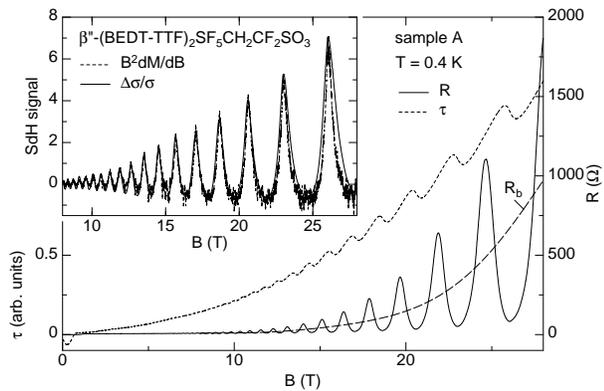,clip=,width=8cm}}
\caption[randm]{Field dependence of the simultaneously measured
torque signal (dotted line) and resistivity (solid line) of sample
$A$. The dashed line is the estimated background magnetoresistance,
$R_b$, resulting in the SdH signal $\Delta \sigma/\sigma$ which agrees
with the derivative of the dHvA signal times $B^2$ (inset).}
\label{randm}
\end{figure}

The $\beta^{''}$-(BEDT-TTF)$_2$SF$_5$CH$_2$CF$_2$SO$_3$
single crystals (labelled $A$ to $C$ hereafter) were grown
by electrocrystallization \cite{schof}. For the transport measurements
15~$\mu$m gold-wire current leads were glued with graphite paste
to the samples. $R$ was measured either with a lock-in amplifier or by
use of a four-point low-frequency ac-resistance bridge with currents of
a few $\mu$A. The aniso\-tropic magnetization, $M$, was measured by means
of a capacitance cantilever torque magnetometer. Thereby, the contacted
crystal $A$ was placed on top of the cantilever plate allowing the
simultaneous measurement of both SdH and dHvA signals. The measurements
were performed at the High Magnetic Field Laboratories in Grenoble in
fields up to 28~T and in Tallahasse up to 30~T.

In first high-field SdH measurements, $R_b$ was estimated
by a polynomial fit to the measured $R$ data \cite{wos99}.
Thereby the oscillation amplitude of $\Delta \sigma/\sigma$
reduced towards lower temperatures for fields above about 20~T
(see Fig.~3 in \cite{wos99}). This is contrary to the standard
theories for magnetic quantum oscillations \cite{tan75}. In order to
get a better estimate for $R_b$, we simultaneously measured the
SdH and dHvA effect of a new and better-quality sample ($A$).

\begin{figure}
  \centerline{\psfig{file=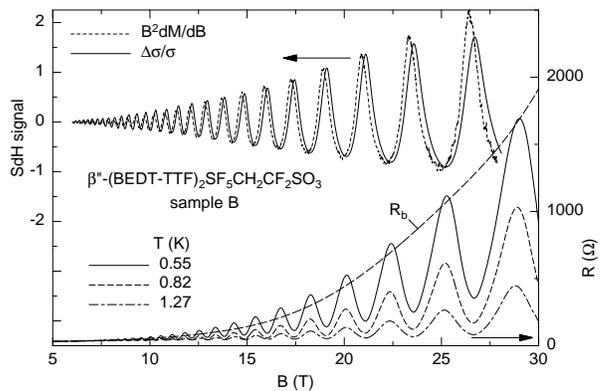,clip=,width=8cm}}
\caption[sdhva]{Magnetoresistance of sample $B$ for different
temperatures. The dashed line shows $R_b$ for $T = 0.55$~K resulting
in the SdH signal $\Delta \sigma/\sigma$ (solid line in the upper part)
which compares well with the derivative of the independently measured
dHvA signal times $B^2$.}
\label{sdhva}
\end{figure}

Figure \ref{randm} shows the torque signal, $\tau$, (dotted line) and
the magnetoresistance (solid line) measured during falling field
which was tilted by about $0.4^\circ$ in order to resolve a
non-zero dHvA signal. In line with previous results \cite{wos00}
the dHvA signal shows an ``inverse sawtooth'' wave form - after
subtraction of a quadratic background magnetization - which can be
explained quantitatively by a 2D theory with fixed chemical
potential \cite{wos00,har96}. The highly asymmetric wave form
becomes more apparent in the strong anharmonicities of $B^2 dM/dB$,
i.e., in the derivative of the dHvA signal times $B^2$ (inset of
Fig.\ \ref{randm}) which is expected to be directly proportional
to the SdH signal $\Delta \sigma/\sigma$ \cite{sho84,rem}.
With the assumption that the SdH signal should be consistent with
the dHvA effect and that the theory for SdH oscillations \cite{ada59}
holds even for magnetoresistance oscillations as large as in the
present case, the background magnetoresistance $R_b$ can be estimated.
The SdH signal $\Delta \sigma/\sigma$ is proportional to the oscillating
part of the density of states at the Fermi level $\Delta
N(\epsilon_F)/N_0$, where $N_0$ is the steady density of states.
Relevant for the magnetic quantum oscillations in the present case
is a single 2D hole-like band, which coexists together with open
1D electron-like bands \cite{bec98}. {\it A priori}, both bands
are expected to contribute to the electronic transport. If, therefore,
the conductivities originating from each band exhibit strong
field and temperature dependences, one has to
disentangle the different contributions from the measured $R$.
For the present case, this means to extract the background
resistance for the 2D band.
With $R_b$ as shown in Fig.\ \ref{randm} (dashed line), a SdH signal
$\Delta \sigma/\sigma $ is obtained which reasonably well agrees
with the thermodynamic quantity $B^2 dM/dB$ (see inset).
For this sample and sample $B$ (see below) a simple polynomial fit
through the data would yield a smaller $R_b$ and, consequently, a too
low $\Delta \sigma/\sigma$ at higher fields.

The estimated $R_b$ for sample $A$ is well inside the range of the
resistance-oscillation amplitude. This is different for samples which
reveal smaller-amplitude oscillations, i.e., which are of less-high quality.
$R_b$ for sample $B$ has to be set at about the maxima of the measured
$R$ in order to reproduce in $\Delta \sigma/\sigma$ approximately the
field dependence of $B^2 dM/dB$ (Fig.\ \ref{sdhva}). For this sample
$R$ was measured in Tallahassee and subsequently $\tau$
was measured in Grenoble (sample 2 in Ref.\ \onlinecite{wos00}),
which explains the slight phase shift.
It is obvious that without the knowledge of the dHvA signal the
chosen $R_b$ seems to be rather arbitrary. For the usual polynomial
estimate of $R_b$,
the $B$ and $T$ dependence of the resulting SdH signal would
contradict the LK theory (see Refs.\ \onlinecite{wos99,zuo99}).
Although there remains an uncertainty in $R_b$ especially towards higher
$B$, the comparison of the magnetic quantum oscillations extracted
from thermodynamic and transport data indeed allows a reliable
estimate of the magnetoresistance.

\begin{figure}
  \psfig{file=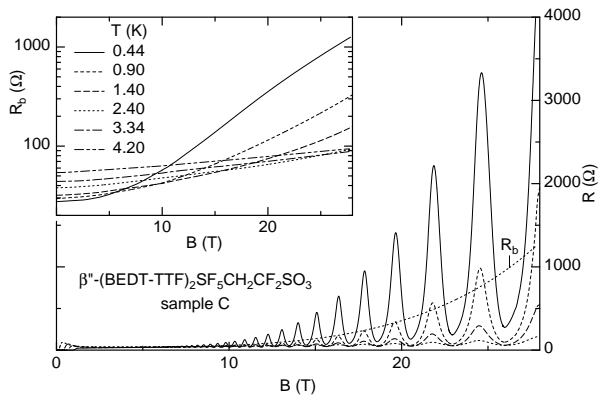,clip=,width=8cm}
\caption[]{Magnetoresistance of sample $C$ for different temperatures.
The dotted line shows $R_b$ at $T = 0.44$~K which is also shown in
a semilogarithmic scale in the inset together with $R_b$ for other
temperatures.}
\label{samplec}
\end{figure}

Finally, Fig.\ \ref{samplec} shows the magnetoresistance of a
high-quality sample ($C$) with very large oscillation amplitudes
in $R$ (three times larger than for sample $B$).
Here, a polynomial fit for $R_b$ results in a reasonable field
and temperature dependence of $\Delta \sigma/\sigma$. The inset of
Fig.\ \ref{samplec} shows the field dependence of $R_b$ for
different temperatures. While for low fields $R_b$
decreases with decreasing $T$ in a metallic-like fashion, insulating
behavior at high fields and low temperatures with $dR/dT < 0$ is
observed. This suggests a metal-insulator transition as $T \rightarrow
0$. At high fields, $R_b$ grows approximately exponentially with
field. This field-induced metal-insulator transition, recognized
qualitatively already earlier \cite{zuo99},
is a unique feature of the present organic metal.

\begin{figure}
  \psfig{file=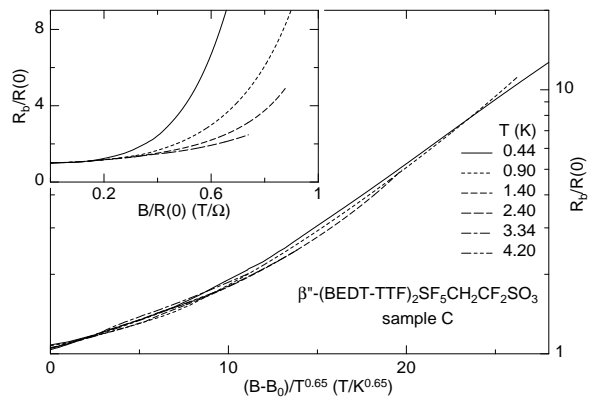,clip=,width=8cm}
\caption[]{Scaling plot of the data shown in the inset of
Fig.\ \ref{samplec}. For $B_0 = 3.5$~T and an exponent
$\alpha = 0.65$ the data collapse on a single curve. A Kohler plot
of the same data (inset) reveals the significant deviation from
the semiclassical theory.}
\label{scalkohl}
\end{figure}

It is important to note that the magnetoresistance cannot be
explained by a conventional semiclassical theory. For
charge carriers with a constant $\tau_s$ on the whole
Fermi surface, the Boltzmann equation predicts that Kohler's
rule is obeyed. However, a Kohler plot of
$R_b/R(0)$ vs $B/R(0)$ for different temperatures (inset of
Fig.\ \ref{scalkohl}), where $R(0)$ for $B=0$ is extrapolated from
$R_b$ at fields above the superconducting phase transition, shows
the failure of the semiclassical theory. This leads to the question
whether the concept of Bloch states for interlayer transport still
has a meaning for the present material \cite{mck98b} and how the
observed field-induced metal-insulator transition and the interlayer
transport in general can be understood.

One explanation for an anomalously large magnetoresistance is
based on the existence of a periodic potential in each layer
\cite{yos95}. A magnetic field applied perpendicular to the
layers then converts the in-plane periodic potential into a
periodic potential along the field direction. When the period is
incommensurate with the layer spacing and
when this potential is stronger than the interlayer hopping rate
the electron wave function would become localized. The strength of
the potential increases with field resulting in an increasing
magnetoresistance \cite{yos95}. For the present material, there is
however no indication for the existence of an in-plane periodic
potential, caused e.g.\ by a density wave. Therefore, it is unclear
whether this kind of incoherent transport is present here.

The metal-insulator transition has been intensively studied in a
number of other materials. For these, usually a universal dynamic
scaling relation can be found which describes the resistivity as
a function of the tuning parameter and temperature \cite{waf99}.
Thereby, the tuning parameter controls the quantum phase
transition, i.e., might be the charge-carrier concentration,
pressure, disorder, or the magnetic field. A scaling
variable which has been found to hold well for the
magnetic-field-tuned superconductor-insulator transition of 2D
films is $(B-B_0)/T^\kappa$, where $B_0$ is the critical field
for the metal-insulator transition and $\kappa$ is a composed
critical exponent. \cite{fis90,mas99,mar99} A first attempt to scale
$R_b$ (inset of Fig.\ \ref{samplec}) directly as a function of
$(B-B_0)/T^\kappa$ did not yield a satisfactory result. However,
the data do collapse onto one single curve when normalizing
$R_b$ by its value at $B=0$ (Fig.\ \ref{scalkohl}). Thereby,
a critical field of $B_0 = 3.5$~T and a critical exponent of
$\kappa = 0.65$ was chosen. $R(0)$ lies between 27.5~$\Omega$
for $T = 0.44$~K and 54~$\Omega$ for $T = 4.2$~K. For sample
$B$ an equally good scaling was obtained with $B_0 = 3.5$~T
and $\kappa = 0.7$. The slight deviations of the different curves
towards higher fields may originate from the uncertainty in $R_b$.
The scaling works very well over about two decades in $R_b$
and about one decade in $B$ and $T$.
For non-zero temperatures above $B_0$ a finite conductivity
should be possible only due to hopping processes. Ideally, for
$T \rightarrow 0$ the magnetoresistance for large enough $B$ should
diverge. However, towards lower temperatures the scaling fails and
$R_b$ rather seems to saturate. \cite{zuo99,wos00b}
A similar deviation from scaling at low $T$ observed in
superconducting 2D films was ascribed to the coupling of the
system to a dissipative bath \cite{mas99}.

The field $B_0$ separates the insulating from the metallic
behavior. Usually, the data for $B < B_0$ should scale onto
a second branch when plotting $R_b/R(0)$ vs $|(B-B_0)/T^\kappa|$.
In the present case, where $B_0 = 3.5$~T is approximately
equal to the upper critical field for superconductivity \cite{wan98},
the resistivity below $B_0$ is strongly influenced by the vortex
dynamics in the superconducting state. Therefore, a reliable scaling
for $B < B_0$ is not possible. For the known quantum phase transitions
the exponent $\kappa$ can be written as $1/z\nu$. For the field-tuned
metal-insulator transition of strongly disordered films a dynamical
critical exponent $z = 1$ and $\nu \ge 1$ for the coherence-length
exponent was predicted and found experimentally ($\nu \approx 1.3$)
\cite{fis90,mas99}. However, recent results deviate from that
prediction and suggest a more complex scenario \cite{mar99}.
Although the exponent $\kappa = 0.65(5)$ found here is close to
the result of \cite{fis90,mas99}, the investigated organic system
is very clean and the metal-insulator transition will probably
fall in a different universality class.

In conclusion, we have shown that for
$\beta^{''}$-(BEDT-TTF)$_2$SF$_5$CH$_2$CF$_2$SO$_3$ the apparent
deviation of the SdH oscillations from conventional behavior depends
on the sample quality and relies strongly on the correct
determination of the background magnetoresistance $R_b$,
which can be achieved by a direct comparison with dHvA data.
$R_b$ exhibits a field-induced
metal-insulator transition which semiclassical theory fails to
describe. The magnetoresistance data for different temperatures
can reasonably well be scaled onto a universal curve.
It remains to be checked whether the unusual behavior
of the SdH oscillations observed in other organic
metals \cite{lau95,san96,bal96} might be understood by a similar
scenario.

We thank R.H. McKenzie for enlightening discussions. The work at
Karls\-ruhe was supported by the DFG and the TMR Programme
of the European Community (contract No.\ ERBFMGECT950077).
J.S.Q.\ was supported by Grant No.\ NSF-DMR-10427. We acknowledge
the National Science Foundation, the state of Florida, and
the U.S.\ Dept.\ of Energy for support of the National High Magnetic
Field Laboratory. Work at Argonne National
Laboratory was supported by the U.S.~Dept.\ of Energy (W-31-109-ENG-38).
Work at Portland State University was supported by NSF
(Che-9904316) and the Petroleum Research Fund (ACS-PRF \# 34624-AC7).


\end{document}